\documentclass[aps,nofootinbib,amsfonts,superscriptaddress]{revtex4}

\usepackage{amsfonts,amssymb,amscd,amsmath}
\usepackage{graphicx}
\usepackage{subfigure}
\usepackage{xcolor}
\usepackage{bm}

\graphicspath{{figures/}}

\newcommand{\dd}{\, \mathrm{d}}

\begin{document}

\title{Isolated photon production and pion-photon correlations\\ 
in high-energy $pp$ and $pA$ collisions}

\author{Victor~P.~Goncalves}
\email{barros@ufpel.edu.br}
\affiliation{High and Medium Energy Group, Instituto de F\'{\i}sica e Matem\'atica, 
Universidade Federal de Pelotas, Pelotas, RS, 96010-900, Brazil} 

\author{Yuri~Lima}
\affiliation{High and Medium Energy Group, Instituto de F\'{\i}sica e Matem\'atica, 
Universidade Federal de Pelotas, Pelotas, RS, 96010-900, Brazil} 

\author{Roman~Pasechnik}
\email{roman.pasechnik@thep.lu.se}
\affiliation{Department of Astronomy and Theoretical Physics, Lund University, 
SE-223 62 Lund, Sweden}
\affiliation{Nuclear Physics Institute of the CAS, 25068 \v{R}e\v{z}, Czech Republic}

\author{Michal~\v{S}umbera}
\email{sumbera@ujf.cas.cz}
\affiliation{Nuclear Physics Institute of the CAS, 25068 \v{R}e\v{z}, Czech Republic}

\begin{abstract}
A phenomenological study of the isolated photon production in high energy $pp$ and $pA$ collisions at RHIC and LHC energies 
is performed. Using the color dipole approach we investigate the production cross section differential in the transverse 
momentum of the photon considering three different phenomenological models for the universal dipole cross section. 
We also present the predictions for the rapidity dependence of the ratio of $pA$ to $pp$ cross sections.
As a further test of the formalism, for different energies and photon rapidites we analyse the correlation 
function in azimuthal angle $\Delta\phi$ between the photon and a forward pion. The characteristic double-peak structure 
of the correlation function around $\Delta \phi\simeq \pi$ observed previously for Drell-Yan pair production is found 
for isolated photon emitted into the forward rapidity region which can be tested by future experiments.
\end{abstract}

\maketitle

\section{Introduction}
\label{Sec:Introduction}

The isolated (prompt) photon production in $pp$ and $pA$ high-energy collisions represents an attractive and 
clean probe for strong interactions in soft \cite{Pasechnik:2016wkt, Acharya:2018dqe, David:2019wpt}
and perturbative  regimes of Quantum Chromodynamics (QCD) \cite{Berger:1990es, Gordon:1994ut,Frixione:1998jh} 
as well as nuclear effects and medium-induced QCD phenomena~\cite{Gelis:2002ki, Peng, Ducloue:2017kkq}. This becomes 
possible due to the absence of QCD-induced final-state interactions associated with absorptive phenomena as well 
as of an energy loss which is in variance to the di-hadron production channels where the final-state absorptive 
corrections are typically very large. The prompt photon production in hadron-hadron and hadron-nucleus collisions can be employed 
to set further constraints on parton density functions (PDFs) in specific kinematic domains not sufficiently well 
explored by HERA \cite{dEnterria:2012kvo,Benic:2018hvb,Schmidt:2015zda}. For this purpose, such studies are also in the focus of ongoing and planned measurements at the LHC \cite{Acharya:2018dqe, Aad:2010sp, 
Khachatryan:2010fm, Zhang:2017xpv, Adam:2015lda} and at RHIC \cite{Adler:2006yt, Adare:2010yw,Adare:2018wgc,David:2019wpt,Yang:2019bjr,Roland:2019cwl}.

At very low-$x$, for example, the primordial transverse momentum evolution of incoming partons and non-linear 
QCD effects such as gluon saturation start to play a significant role whose reliable first-principle analysis 
represents a long-standing theoretical challenge. In the case of high-energy $pA$ collisions, main issues 
concern a proper description of initial/final state effects in 
multiple interactions with a nuclear target. Another widely discussed problem is associated with propagation 
of partons in the nuclear environment. Such processes, as the Drell-Yan (DY) pair production, studied recently 
by some of the authors in Refs.~\cite{Basso:2015pba,BGKNP,Goncalves:2016qku}, as well as the isolated 
photon production at high-$p_T$, provide efficient means for phenomenological analysis of various 
nuclear effects such as the nuclear shadowing and initial-state interactions determined by saturation \cite{salgado}.

\begin{figure}[t]
 \centering
 \subfigure[]{
 \scalebox{0.33}{\includegraphics{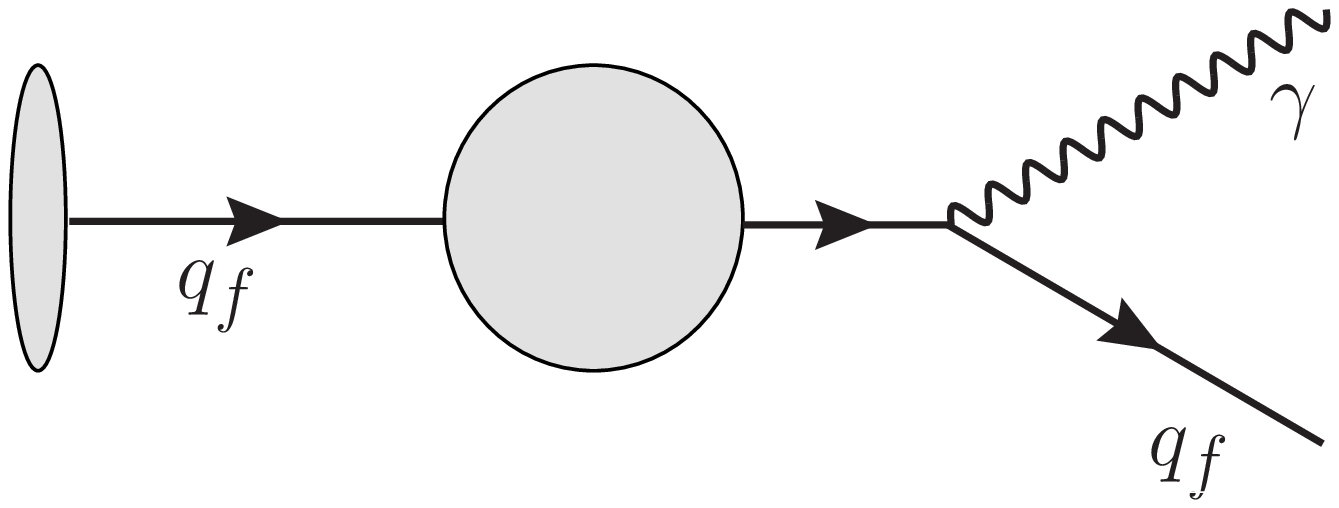}}
 \label{fig:gb_dir}
 }
 \centering
 \subfigure[]{
 \scalebox{0.33}{\includegraphics{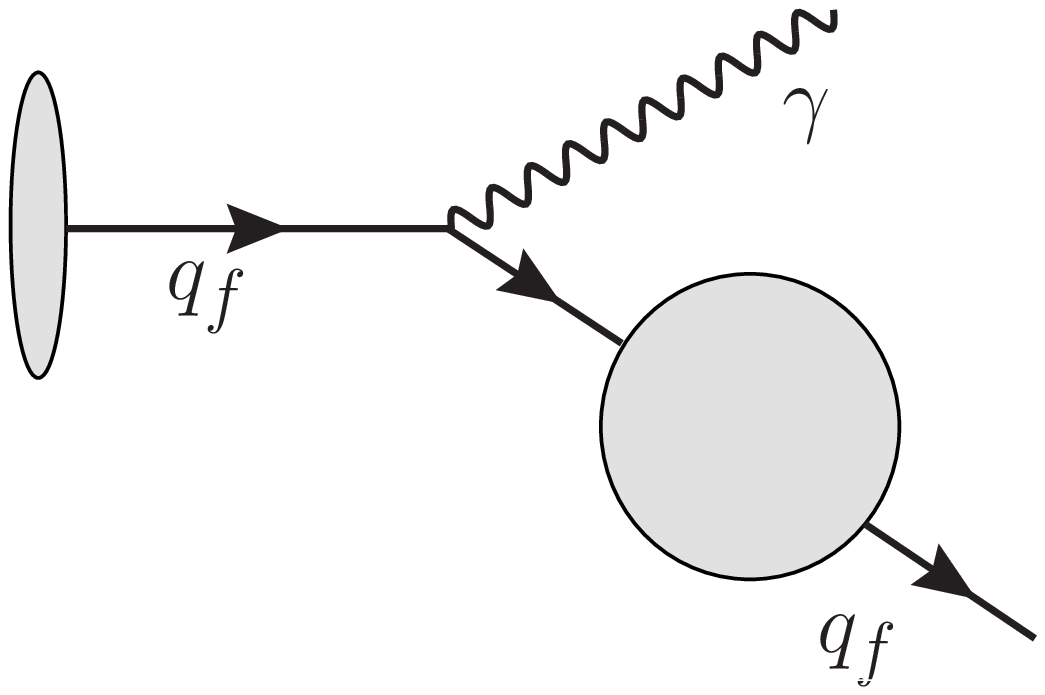}}
 \label{fig:gb_frag}
 }
 \centering
 \subfigure[]{
 \scalebox{0.33}{\includegraphics{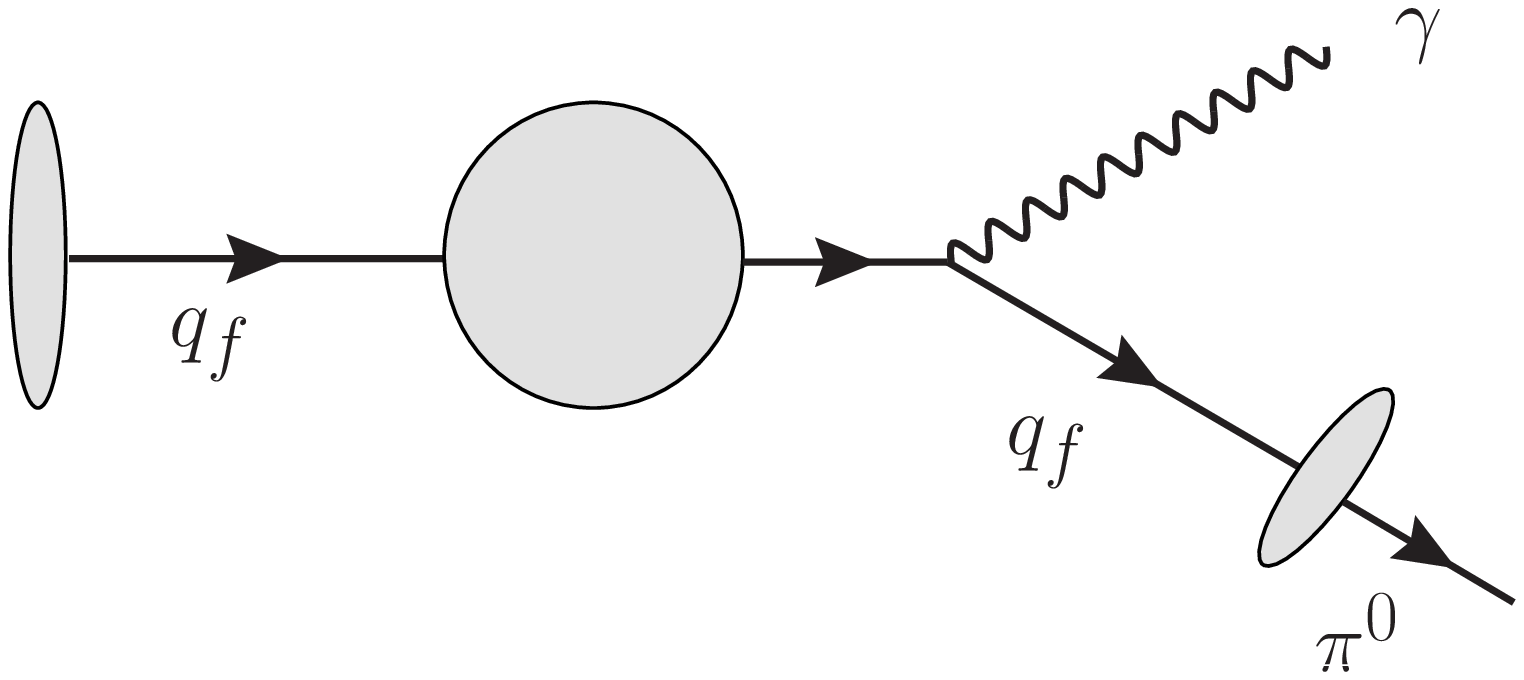}}
 \label{fig:gb_frag1}
 }
 \caption{Diagrams (a) and (b) illustrate the $\gamma$ 
 Bremsstrahlung process off a projectile a quark (antiquark) 
 of flavour $f$ either after and before its interaction with 
 the color field of the target (denoted by a shaded circle), 
 respectively. Diagram (c) corresponds to one of the possible
 contributions to the pion-photon pair production in the color 
 dipole picture.}
 \label{fig:gb_dip}
 \end{figure}

In this paper, we investigate the isolated photon production off the proton and nuclear targets in low-$x$ regime 
of QCD in the framework of the phenomenological color dipole formalism (see
e.g.~Refs.~\cite{zkl,nik,bhq97,kst99,krt01,dynuc,rauf}). In the dipole picture, the real
photon production is considered as $\gamma$ Bremsstrahlung off a fast projectile quark propagating 
through the low-$x$ color field of the target \cite{kst99} as illustrated in Fig.~\ref{fig:gb_dip} (panels (a) and (b)).
In this case, the photon radiation occurs both after and before the quark scatters off the target 
and the corresponding amplitudes interfere. As a result of such interference, the photon Bremsstrahlung 
process can be viewed as scattering of a $q\bar q$ dipole with a given transverse separation. This 
in variance to the conventional parton model where the same process in the center-of-mass frame 
is given by the Compton scattering. The difference between both descriptions illustrates the well known 
fact that although cross sections are Lorentz invariant, the partonic interpretation of the corresponding 
processes depends on the reference frame.

The key ingredients of the dipole formula for the differential cross section 
of the considered process are the light-cone (LC) wave function of the initial state describing the real 
photon radiation off the projectile quark as well as the universal dipole-target cross section related 
to the dipole $S$ matrix, $\sigma_{q\bar q}(x,\rho)$, which can be determined phenomenologically, for example, 
by a fit to the  Deep Inelastic Scattering (DIS) data at low-$x$ \cite{GBW} or to a Drell-Yan $pp$ data of 
good quality. At small Bjorken $x$ (or at high energies), 
the universality of the dipole cross section stems from the fact that color dipoles in QCD are the eigenstates 
of interaction  with a  fixed transverse separation, $\rho$ \cite{zkl}. 

Remarkably, since the lifetime of partonic fluctuation in the laboratory frame 
is enhanced by a factor $\sqrt{s}/m_p$ wrt to the lifetime in the centre-of-mass system, the phenomenological dipole approach appears to effectively take into account the higher-order 
QCD corrections. For example, it provides the predictions for the DY process at the same level of precision 
as the Next-to-Leading-Order (NLO) collinear factorisation framework \cite{rauf}. Besides, as a consequence 
of universality, the dipole formulation provides a unified description of a variety of inclusive and 
diffractive observables of particle production processes in lepton-hadron, hadron-hadron, hadron-nucleus 
and nucleus-nucleus collisions at high energies (for particular examples, see e.g.~Refs.~\cite{nik,nik_dif,bhq97,
kst99,krt01,npz,pkp}). In the high-energy limit, the projectile quark effectively probes dense gluonic field 
in the target with the dipole cross section effectively accounting for the non-linear effects due 
to multiple scatterings.

The goal of the current work is the following. First, we update the previous studies presenting predictions 
for the transverse momentum distribution of isolated photons produced at the RHIC and LHC energies. 
Moreover, we also make predictions for the ratio between the proton-lead ($pPb$) and proton-proton ($pp$) 
cross sections at the LHC for different values of the photon (pseudo-)rapidity. Second, we present a detailed analysis 
of the azimuthal correlation between the photon and a pion that emerges from a projectile quark hadronisation 
at forward rapidities\footnote{Similar correlations in di-hadron, real photon-hadron and dilepton-hadron channels
have been previously reported in Refs.~\cite{Marquet,stasto,stastody,Stasto:2018rci,amir}.} 
(see Fig.~\ref{fig:gb_dip} (c)). In this paper, we present new results for such an observable 
for $pp$ collisions at RHIC ($\sqrt{s}=500$ GeV) and LHC ($\sqrt{s}=14$ TeV), as well for $pAu$ collisions 
at RHIC ($\sqrt{s}=200$ GeV) and $pPb$ collisions at the LHC ($\sqrt{s}=8.8$ TeV). In order to estimate 
the related theoretical uncertainties in our predictions, we consider three different approaches  
to saturation effects \cite{GBW,hdqcd,aamqs}.

The paper is organized as follows. In Sect.~\ref{Sec:formalism}, a brief overview of isolated photon 
production in the color dipole framework is provided. In Sect.~\ref{Sec:results}, we present our numerical
results for the transverse momentum distributions of the produced isolated photon as well as 
the $pA$-to-$pp$ ratio of the production cross sections. Furthermore, the pion-photon azimuthal correlation 
function is evaluated for $pp$ and $pA$ collisions at the characteristic RHIC and LHC energies for different 
photon and pion rapidities. Finally, in Sect.~\ref{Sec:summary}, our main conclusions are summarized.

\section{Color dipole picture of real photon Bremsstrahlung}
\label{Sec:formalism}

Consider first the isolated photon production in $pp$ collisions in the target rest frame. In the high energy limit, each of the first two diagrams (a) and (b) in Fig.~\ref{fig:gb_dip} 
in the impact-parameter space can be represented as a convolution of the LC wave function of the projectile quark $|q\rangle$ fluctuation into its lowest $|q\gamma\rangle$ Fock state and a scattering amplitude of a quark off the target $T$ at a given impact parameter \cite{bhq97,kst99}. 
Here, $T$ denotes either the proton $p$ or a nucleus target with an atomic mass $A$.

In what follows, we work in terms of usual LC
(longitudinal) momentum fractions of the isolated photon, 
$x_1$ and $x_2$, taken from the incoming proton momenta 
$p_1$ and $p_2$, respectively, such that
\begin{eqnarray}
x_1=\frac{p_\gamma^+}{p_1^+} = {\frac{p_T}{\sqrt{s}}}\,e^{\eta}\,, \qquad
x_2=\frac{p_\gamma^-}{p_2^-} = {\frac{p_T}{\sqrt{s}}}\,e^{-\eta}\,, \qquad
x_1-x_2\equiv x_F\,,
\end{eqnarray}
where $p_T$, $\eta$ and $x_F$ are the transverse momentum, pseudorapidity
and the Feynman variable of the photon. The initial-state quark $|q\rangle$ 
and the final-state quark accompanied by a Weiz\"acker-Williams 
photon, $|q\gamma\rangle$, propagate at different impact
parameters. Indeed, due to the $\gamma$ Bremsstrahlung
the final quark gets a transverse shift with respect 
to the initial one, $\Delta{\bf r} = \alpha \bm{\rho}$, where 
$\alpha$ is the fractional LC momentum taken by 
the radiated photon off the projectile quark and $\bm{\rho}$ 
is the quark-$\gamma$ transverse separation.

The amplitudes (a) and (b) in Fig.~\ref{fig:gb_dip} corresponding to scattering of $|q\rangle$ and $|q\gamma\rangle$ Fock states off the target $T$, respectively, interfere. As a result, the matrix element squared for the isolated photon production integrated over the impact parameter of the initial quark is expressed in terms of the universal $q\bar q$ dipole-target cross section $\sigma^T_{q\bar q}(\Delta{\bf r},x)$ as a function of the transverse separation $\Delta{\bf r}$ and the standard Bjorken variable of the process $x$ which is taken to be equal to $x_2$ in what follow. The cross section for the real photon production differential in photon transverse momentum $p_T$ and pseudorapidity $\eta$,
\begin{eqnarray} 
\label{eq:gb_cs}
\frac{d \sigma (p \,T\rightarrow {\gamma} X)}{\dd^2 p_T d\eta} = \frac{2 p_T}{\sqrt{s}}\,\cosh(\eta)\,\frac{x_1}{x_1 + x_2}\,
\sum_f  \int_{x_1}^1 \frac{d \alpha}{\alpha^2} \left[ q_f(x_1/\alpha,\mu_F^2) + \bar{q}_{{f}}(x_1/\alpha,\mu_F^2)  \right] 
\frac{d\sigma^f (q \,T \rightarrow q{\gamma}X)}{d\ln \alpha d^2p_T}
\end{eqnarray}
is typically found in terms of the unpolarised projectile quark (antiquark) 
collinear PDFs $q_f$ ($\bar{q}_f$) corresponding to (valence and sea) 
flavor $f=u,d,s,c$ as functions of the momentum fraction of the projectile 
quark taken from the parent nucleon $x_q = x_1/\alpha$ and the QCD factorisation 
scale $\mu_F=p_T\equiv |{\bf p}_T|$. The differential cross section of 
the high-$p_T$ real photon production in the quark-target scattering 
subprocess is represented in the dipole picture as
\begin{eqnarray}
\label{ptdistcc}
\frac{d\sigma^f (q \,T \rightarrow q{\gamma}X)}{d \ln \alpha d^2p_T} 
& = & \frac{1}{(2\pi)^2}\, 
\int d^2\rho_1 \dd^2\rho_2 \exp[i{\bf p}_T \cdot (\bm{\rho}_1 - \bm{\rho}_2)]\, \Psi(\alpha,\bm{\rho}_1,m_f) 
\Psi^{*}(\alpha,\bm{\rho}_2,m_f) \nonumber \\
 & \times & \frac{1}{2}\left[ \sigma^T_{q\bar q}(\alpha \bm{\rho}_1,x_2) 
 + \sigma^T_{q\bar q}(\alpha \bm{\rho}_2,x_2) - 
 \sigma^T_{q\bar q}(\alpha|\bm{\rho}_1 - \bm{\rho}_2|,x_2)\right]\,, 
\end{eqnarray}
where $m_f$ is the constituent quark mass, and $\Psi(\alpha,\bm{\rho},m_f)$ 
is the LC wave function of the real photon radiation off a quark 
with flavor $f$. Following Ref.~\cite{GBW}, we take the constituent 
quark mass values to be $m_u = m_d = m_s = 0.14$ GeV and $m_c = 1.4$ GeV
in our numerical analysis below. For the cross section differential in photon 
${\bf p}_T$ the quark-$\gamma$ transverse separations amplitude 
and its conjugated are considered to be different and are denoted 
as $\bm{\rho}_{1,2}$. In this case, the overlap of the photon 
Bremsstrahlung wave functions in Eq.~(\ref{ptdistcc}), summed over 
the transverse polarisations of the radiated hard photon, reads
\begin{eqnarray} 
\label{Psi2}
\Psi(\alpha,\bm{\rho}_1,m_f) \Psi^{*}(\alpha,\bm{\rho}_2,m_f) 
= \frac{\alpha_{\rm em}e_f^2}{2 \pi^2} \Bigg\{
m_f^2 \alpha^4 {\rm K}_0\left(\tau \rho_1\right)
{\rm K}_0\left(\tau \rho_2\right) + \left[1+\left(1-\alpha\right)^2\right]\tau^2
\frac{\bm{\rho}_1\cdot\bm{\rho}_2}{\rho_1 \rho_2}
{\rm K}_1\left(\tau \rho_1\right)
{\rm K}_1\left(\tau \rho_2\right)
\Bigg\}\,, 
\end{eqnarray}
where $\bar\alpha\equiv 1-\alpha$, 
$\alpha_{\rm em}$ is the fine structure constant, $e_f$ 
is the charge of the projectile quark, $\rho_{1,2}\equiv|\bm{\rho}_{1,2}|$, 
$\tau =  \alpha m_f$ and the modified Bessel functions of the second kind 
are denoted as ${\rm K}_{0,1}$. In fact, the photon transverse momentum
provides a hard scale for the considering process which ensures the validity
of the perturbative approximation which has been used in the computation of 
the photon wave function in Eq.~(\ref{Psi2}).

We would like to analyse the correlation in the azimuthal angle between
the final-state photon and a hadron emerging due to hadronisation of the
projectile (anti)quark associated with the photon radiation. An analogous 
analysis for the DY process with deeply-virtual photon has been performed 
earlier in the impact parameter representation in Ref.~\cite{dynuc}, although
the corresponding numerical analysis is very challenging. More recently, in Ref.~\cite{Basso:2015pba} a numerical calculation of the differential DY 
cross section derived in Refs.~\cite{jamal,amir,Dominguezetal12,tese} has been 
performed directly in momentum representation. We adopt the same formalism
for the considering case of real high-$p_T$ photon production in association 
with the leading hadron $h$, namely,
\begin{eqnarray}
\frac{d \sigma(p\,T \to h {\gamma} X)}{d \eta d y_h d^2p_T d^2p_T^h } 
& = & 
\frac{\alpha_{\rm em}}{2\pi^2}
\int_{\frac{x_h}{1-x_1}}^1 \frac{\dd z_h}{z_h^2} 
\sum_f  e_f^2 D_{h/f}(z_h,\mu_F^2)\, 
x_p q_f(x_p,\mu_F)\, S_{\perp}\, F_T(x_g, k^g_T)\, 
\frac{\bar z z^2(1+\bar z^2){k^g_T}^2}
{P_T^2 ({\bf P}_T + z{\bf k}^g_T )^2}\,,
\label{eq:dy-dijet}
\end{eqnarray}
where the key kinematical variables are determined 
as follows
\begin{eqnarray}
&& 
x_h \simeq \frac{p_T^h}{\sqrt{s}}\,e^{y_h} \,, \quad 
x_p = x_1 + \frac{x_h}{z_h} \,, \quad 
z = \frac{x_1}{x_p} \,, \quad
x_g = x_1\,e^{-2\eta} + \frac{x_h}{z_h}\,e^{-2y_h} \,,
\\ 
&& 
{\bf k}^g_T = {\bf p}_T + {\bf k}^q_T \,, \qquad
{\bf P}_T = \bar z{\bf p}_T - z{\bf k}^q_T \,, \qquad
{\bf k}^q_T = \frac{{\bf p}_T^h}{z_h} \,,
\end{eqnarray}
Here, for simplicity, we are considering the light quark flavors 
$f=u,d,s$ only and neglect terms proportional to $m_f$ due 
to $p_T\gg m_f$. In Eq.~(\ref{eq:dy-dijet}), $D_{h/f}$ stands 
for the fragmentation function of the projectile quark $q_f$ 
(which has emitted the photon) into a final-state (light) 
hadron $h$ carrying the transverse momentum ${\bf p}_T^h$ 
that is supposed to be detected in a measurement. The remaining 
kinematic variables are defined as follows: $y_h$ is 
the rapidity of the hadron $h$ in the final 
state, respectively, $z_h$ and $x_h$ are the LC momentum 
fraction taken by the hadron $h$ from the parent quark 
$q_f$ and the incoming proton, ${\bf P}_T$ 
is the relative transverse momentum between the photon and 
the quark $q_f$, ${\bf k}^q_T$ is the transverse
momentum of the projectile quark $q$ (before it fragments 
into a hadron $h$), ${\bf k}^g_T$ is the transverse momentum 
of the exchanged gluon in the $t$-channel. Finally, $S_{\perp}$ 
denotes the transverse area of the considered target $T$ whose 
explicit form is irrelevant for our purposes here, $F_T(x_g, k^g_T)$ 
represents the so-called unintegrated gluon distribution function 
(UGDF) in the target $T$. In the saturation regime and for the soft
gluon $k^g_T$, the latter can be found in terms of a Fourier 
transform of the dipole cross section $\sigma^T_{q\bar q}$
\cite{stastody}. Note, the momentum fractions $z$ and $x_p$ share
the same physical meaning as $\alpha$ and $x_q$ introduced above
in Eq.~(\ref{eq:gb_cs}), respectively. A different notation is
used here since $z$ and $x_p$ are now related the hadron kinematic
variables $z_h$, $y_h$ and $p_T^h$ in the final state.

One of the important observables sensitive to the dynamics of 
saturation is the correlation function $C(\Delta \phi)$ in azimuthal
angle $\Delta \phi$ between the final state photon and hadron 
(for more details, see e.g.~Ref.~\cite{Basso:2015pba}). Assuming 
the isolated photon to be a trigger particle, the correlation 
function can be built as follows
\begin{eqnarray}
C(\Delta \phi) = \frac{ 2\pi\, \int_{p_T, p_T^h > p_T^{\rm cut}} 
dp_T p_T \; dp_T^h p_T^h \; 
\frac{d \sigma(p \,T \to h \gamma X)}{d \eta d y_h d^2p_T d^2p_T^h }}
{\int_{p_T > p_T^{\rm cut}} dp_T p_T \; 
\frac{d \sigma(p \,T \rightarrow \gamma X)}{d\eta d^2 p_T} }\,,
\label{corr}
\end{eqnarray}
in terms of the low cut-off $p_T^{\rm cut}$ on transverse momenta 
of the resolved $\gamma$ and $h$. In the denominator, we have 
the cross section for inclusive photon production. For consistency,
the latter can be straightforwardly obtained by integrating 
photon-hadron cross section in Eq.~(\ref{eq:dy-dijet}) over 
the hadron momentum and rapidity as well as over $\Delta \phi$.
This way, one arrives at the following expression
\begin{eqnarray}
\frac{d \sigma(p\,T\rightarrow \gamma X)}{d\eta d^2 p_T} 
& = & \frac{\alpha_{\rm em}}{2\pi^2}
\int_{x_1}^1 \frac{dz}{z} 
\int d^2 k^g_T \sum_f  e_f^2 x_p q_f(x_p,\mu_F)\, S_{\perp}\, 
F_T(x_g, k^g_T) 
\frac{z^2(1 + {\bar z}^2){k^g_T}^2}{ p_T^2 
({\bf p}_T - z{\bf k}^g_T)^2} \,.
\label{eq:dy-boson}
\end{eqnarray}

For the numerical analysis of the isolated photon observables we need 
to specify a reliable parametrization for the dipole cross section
\cite{zkl}, $\sigma^T_{q\bar q}(r,x)$. The latter contains 
an important information about possible non-linear QCD (or saturation) 
effects in the hadronic state (see for a detailed discussion of saturation 
phenomena, e.g. Ref.~\cite{hdqcd}). In the case of $pp$ collisions, we 
should specify the universal dipole cross section off the proton target. 
Due to universality of dipoles as eigenstates of interaction in QCD, 
such a quantity is typically obtained from a phenomenological analysis 
of the precision data on DIS available from the HERA collider.
For comparison with previous results existing in the literature, we 
traditionally consider the phenomenologically very successful 
Golec-Biernat--Wusthoff (GBW) model \citep{GBW} relying on 
a simple saturated ansatz
\begin{equation}
\label{gbw}
\sigma^p_{q\bar q}(r,x) = \sigma_0
\left(1 - e^{-\frac{r^2 Q_{s,p}^2(x)}{4} } \right)\,,
\end{equation}
with the proton saturation scale
\begin{equation}
Q_{s,p}^2(x) = Q_0^2\left( \frac{x_0}{x} \right)^\lambda \,,
\label{satsca}
\end{equation}
where the model parameters $Q_0^2 = 1$ GeV${}^2$, 
$x_0 = 3.04\times10^{-4}$, $\lambda = 0.288$ and 
$\sigma_0 = 23.03 $ mb were obtained from the fit 
to the DIS data. 
Besides, we consider the solution of the Balitsky-Kovchegov 
equation \cite{BAL,kov} with running coupling obtained 
in Ref.~\cite{aamqs} as an alternative model for the dipole-proton 
cross section, denoted as AAMQS hereafter.
Likewise, its initial conditions were constrained 
by a fit to the HERA DIS data. 
Finally, yet another phenomenological saturation model for 
$\sigma^p_{q\bar q}(r,x)$ based upon the Color Glass Condensate 
(CGC) approach~\cite{iim}
\begin{eqnarray}
\sigma^p_{q\bar q}(r,x) = \sigma_0 \times
\left\{ \begin{array}{ll} {\mathcal N}_0\, 
\left(\frac{ r \, Q_{s,p}}{2}\right)^{2\left(\gamma_s + 
\frac{\ln (2/r Q_{s,p})}{\kappa \,\lambda \,Y}\right)}  
& \mbox{$r Q_{s,p} \le 2$} \\
1 - \exp^{-A\,\ln^2\,(B \, r  Q_{s,p})}   
& \mbox{$r Q_{s,p}  > 2$} 
\end{array} \right.
\label{eq:bcgc}
\end{eqnarray}
has been utilised for comparison and in order to estimate the sensitivity 
of our predictions to dynamics of the saturation effects.
Here, $\kappa = \chi''(\gamma_s)/\chi'(\gamma_s)$, where $\chi$ is 
the LO BFKL characteristic function, and the coefficients $A$ and 
$B$ are uniquely determined from the continuity condition for 
the dipole cross section and its derivative 
with respect to $r\,Q_{s,p}$ at $r\,Q_{s,p}=2$. 

While the dipole cross section off the proton target is well-constrained 
and tested by ample $ep$ and $pp$ phenomenology, in the case of a heavy 
nucleus target the data are not as precise as for the proton one while 
the modelling of the corresponding dipole cross section is still a subject 
of continuous debates. One possible alternative present in several 
studies in the literature is to consider the Glauber-Mueller 
(GM) approach \cite{glauber,mueller90} based upon resummation of 
all the multiple elastic rescattering diagrams for the $q \bar q$ 
dipole propagation through the nucleus target. In this model, 
the dipole-nucleus cross section reads
\begin{eqnarray}
\sigma^A_{q\bar q}(r,x) = 2 \int d^2 b_A 
\left\{  1 - \exp \left[
-\frac{1}{2} \sigma^p_{q\bar q}(r,x)\,
T_{A}(b_A) \right] \right\} \,,
\label{Na_Glauber}
\end{eqnarray}
where $T_{A}(b_A)$ is the nuclear thickness function 
which is typically obtained from the Woods-Saxon distribution 
for the nuclear density normalized to the atomic mass $A$, and $b_A$ is
the impact parameter of the dipole with respect to the nucleus center.
Another possibility is to consider a solution the running-coupling 
Balitsky-Kovchegov (rcBK) equation for the nuclear case discussed 
e.g. in Refs.~\cite{raju_nuc,hl_nuclear}, which takes into account 
mutual interactions of the gluonic ladders exchanged between 
the dipole and the nucleus. These two approaches include different 
diagrams and have distinct predictions for the onset of the saturation 
phenomena. 

In the next Section, we perform a numerical analysis of 
the nuclear modification factor $R_{pA}$ (the $pA$-to-$pp$ ratio 
of the differential cross sections) and the azimuthal correlation 
for isolated photon production and compare predictions obtained 
with these two models for the dipole cross section off the nucleus.
\begin{figure}[t]
\begin{center}
\scalebox{0.45}{\includegraphics{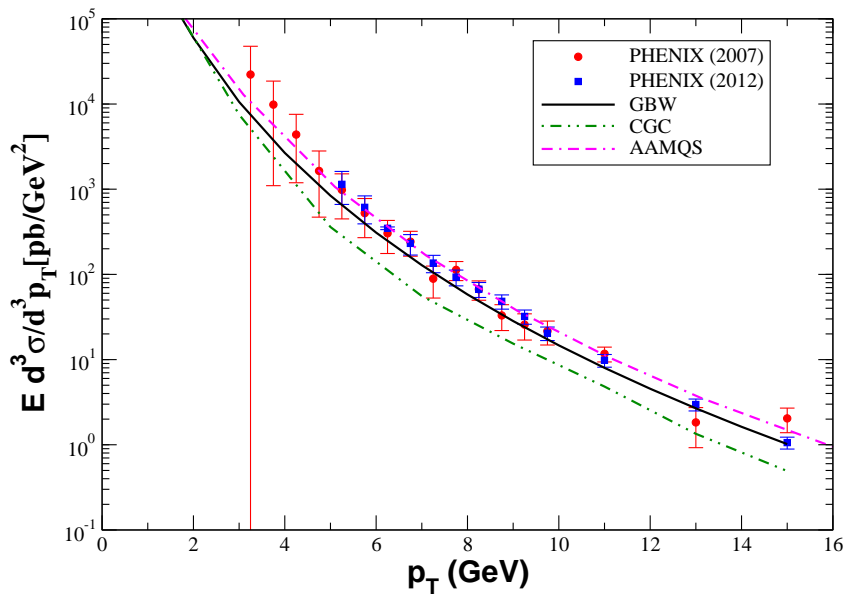}}
\caption{The isolated photon transverse-momentum spectra in $pp$ collisions 
at $\sqrt{s} = 0.2$ TeV and at mid-rapidity, $\eta = 0$, 
obtained using the different models for the dipole cross 
section discussed in the text. The experimental data are 
from the PHENIX experiment~\cite{phenix_photon}.}
\label{fig:Spectra_phenix}
\end{center}
\end{figure}
\begin{figure}[t]
\begin{center}
\scalebox{0.33}{\includegraphics{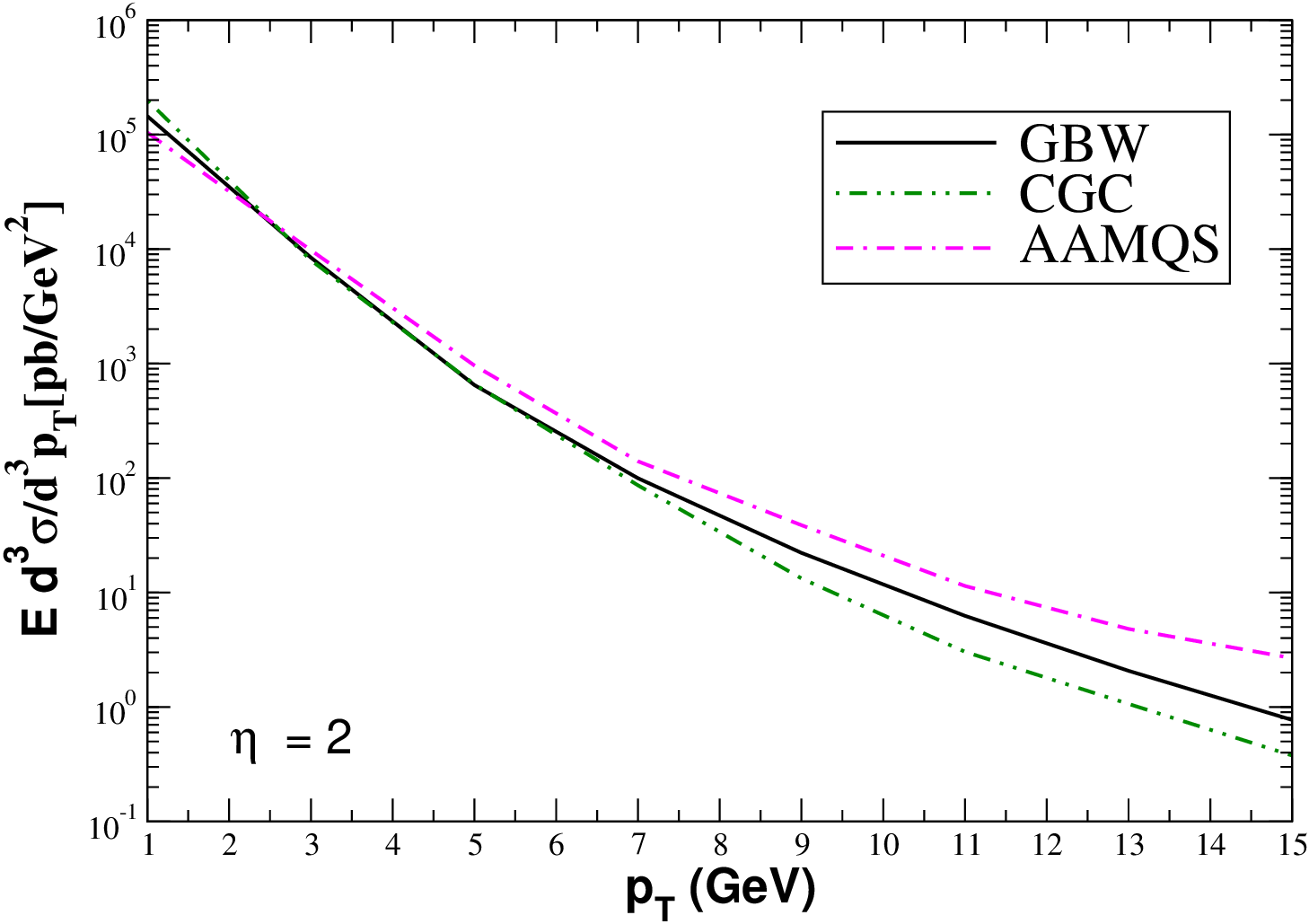}}
\scalebox{0.33}{\includegraphics{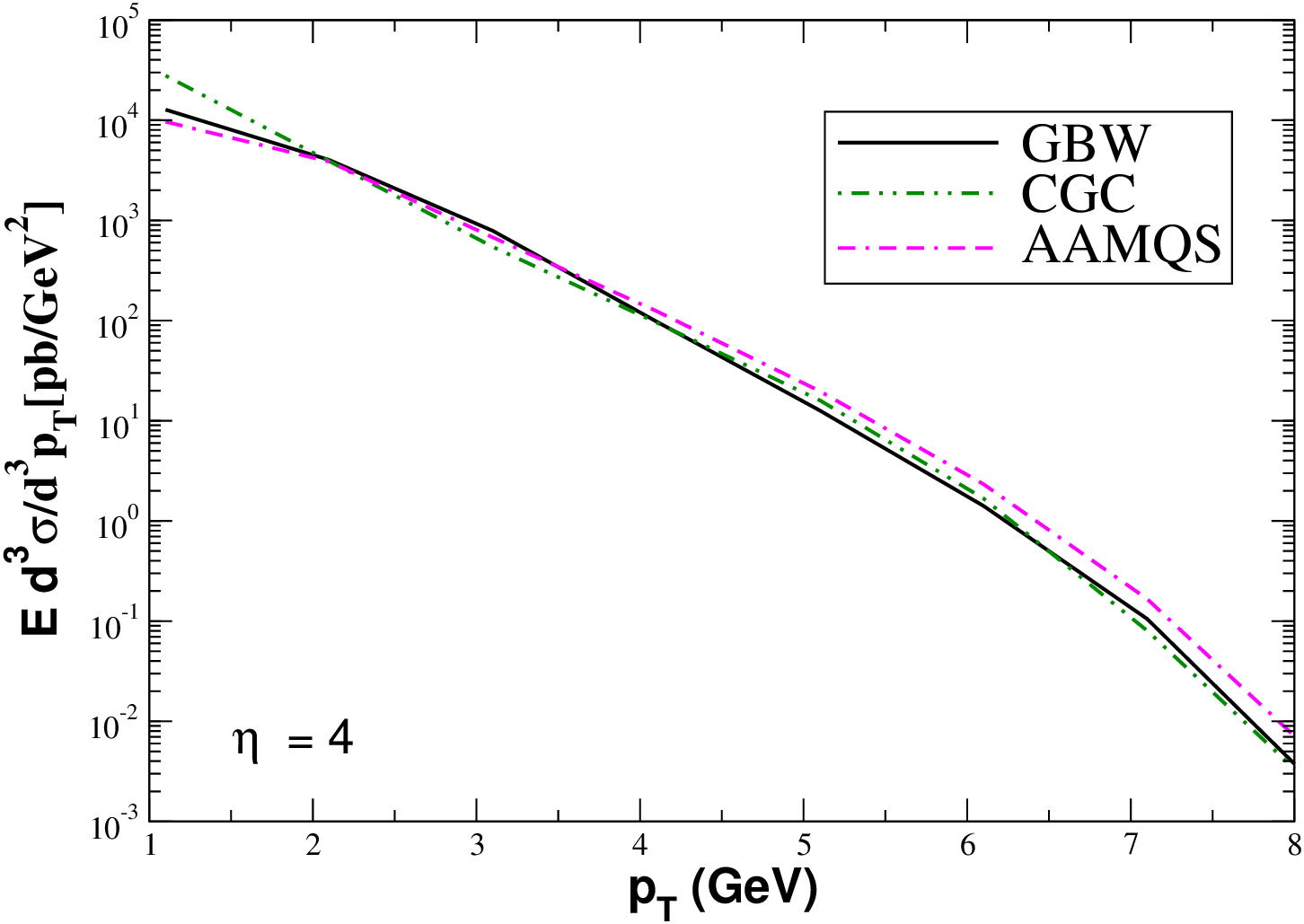}}
\caption{The isolated photon transverse-momentum spectra in $pp$ collisions 
at $\sqrt{s} = 0.5$ TeV of RHIC experiments for two distinct 
values of the photon pseudo-rapidity $\eta$. The results are 
presented for different models for the dipole cross section discussed 
in the text.}
\label{fig:Spectra_rhic}
\end{center}
\end{figure}
\begin{figure}[t]
\begin{center}
\scalebox{0.33}{\includegraphics{spectra_14000_eta4D}}
\scalebox{0.33}{\includegraphics{spectra_14000_eta6D}}
\caption{The isolated photon transverse-momentum spectra in $pp$ collisions 
at $\sqrt{s} = 14$ TeV of the LHC experiments for two distinct 
values of the pseudo-rapidity $\eta$. The results are presented 
for different models for the dipole cross section discussed 
in the text. }
 \label{fig:Spectra_lhc}
 \end{center}
 \end{figure}
\begin{figure}[t]
\begin{center}
\scalebox{0.45}{\includegraphics{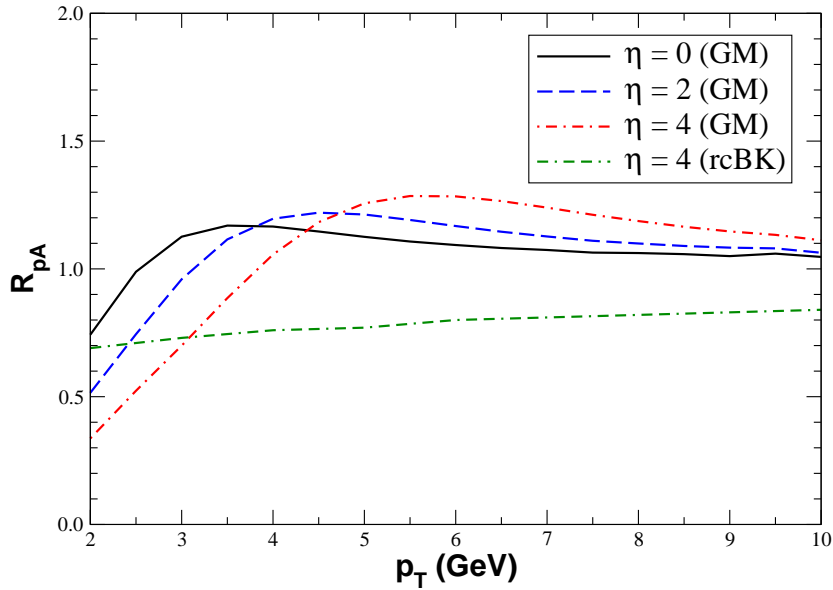}}
\caption{Transverse momentum dependence of the normalized 
nuclear modification factor $R_{pA}$ for isolated photon 
production in proton-lead collisions ($A=208$) at the LHC 
($\sqrt{s} = 8.8$ TeV) for several selected values of the 
photon pseudorapidity $\eta$ and for two distinct (GM and rcBK) 
models of the dipole-nucleus cross 
section. 
}
\label{fig:ratio}
\end{center}
\end{figure}

\section{Numerical results}
\label{Sec:results}

In this Section, we present numerical results for the isolated photon 
production in the $pp\rightarrow \gamma X$ process in the framework 
of color dipole formalism. In this analysis, we employ three phenomenological
parametrizations for the dipole cross section discussed above and
use the CT10 NLO parametrization for the projectile quark PDFs \cite{ct10} 
(both sea and valences quarks are included). 

In Fig.~\ref{fig:Spectra_phenix} we compare our predictions with the PHENIX 
data \cite{phenix_photon} for isolated photon production at mid-rapidity 
in $pp$ collisions at $\sqrt{s} = 0.2$ TeV obtained by using three distinct
models for the dipole cross section off the proton target. We can see that 
the GBW and AAMQS models describes the data quite well while the CGC model
underestimates the data. Note that our results rely on existing
parameterisations for the dipole cross section fitted to the HERA data, 
without any additional free parameters. In particular, no NLO $K$-factor has 
been imposed in the calculations, in contrast to the collinear QCD approach
where such factor is required. 

In Fig.~\ref{fig:Spectra_rhic} we present our predictions for isolated photon
production in $pp$ collisions at $\sqrt{s} = 0.5$ TeV and for two distinct
values for the photon pseudo-rapidity, $\eta=2$ (left panel) and $\eta=4$
(right panel). Here, we have selected forward rapidities in order to probe
small values of $x_2$ in the validity domain of the dipole approach. We
expect that in this case the direct photon $p_T$ spectra are more sensitive
to the treatment of the saturation effects. 

The results presented in Fig.~\ref{fig:Spectra_rhic} (left panel) confirm
that this expectation is valid already for $\eta = 2$. Here, the predictions
for the photon spectrum are similar at small $p_T$'s but start to deviate
significantly at $p_T>6$ GeV. In particular, the AAMQS result, associated to
the solution of the rcBK equation, predicts larger values for the spectra at
large $p_T$'s than those for the GBW and CGC models. 

In contrast, the results for $\eta = 4$ shown in Fig.~\ref{fig:Spectra_rhic}
(right panel) indicate that at such large rapidities one cannot distinguish
the predictions of the different dipole models. Indeed, the dipole approach
becomes more precise for smaller values of $x_2$. In addition, such small
difference between the dipole model predictions is partly due to the fact
that here we probing the photon $p_T$ spectrum in the edge of the phase space
where its behaviour is determined essentially by the kinematics of the
process.

Our predictions for $pp$ collisions at the LHC energy ($\sqrt{s} = 14$ TeV)
and for two different values of the photon pseudo-rapidity are presented in
Fig.~\ref{fig:Spectra_lhc}. Similarly to what was observed at RHIC energies,
we found that the AAMQS prediction yields a higher spectrum than the other
models, particularly, at large photon transverse momenta while the CGC and
GBW parametrizations provide similar predictions. In principle, future
experimental data at large $p_T$ can be used to discriminate between the
AAMQS and GBW models. Note that at small $p_T$, however, the AAMQS prediction
becomes slightly below the GBW one.

In order to estimate the impact of the nuclear effects in the predictions 
for the isolated photon production in proton-lead ($pPb$) collisions 
at the LHC ($\sqrt{s} = 14$ TeV), in Fig.~\ref{fig:ratio} we present 
our predictions for the photon transverse momentum dependence of 
the nuclear modification factor $R_{pA}$ defined as a ratio between 
the nuclear and proton differential cross sections, normalized by 
the atomic mass $A$. The predictions derived using the Glauber-Mueller
approach for the dipole-nucleus cross section, Eq.~(\ref{Na_Glauber}), are
denoted as ``GM'' in the figure. This model predicts that $R_{pA}$ 
becomes smaller than one at small $p_T$ while the nuclear effects become
essentially negligible at large $p_T$. Moreover, the position of 
the maximum depends on the rapidity and shifts towards larger $p_T$'s 
when the rapidity is increased. In contrast, when a solution of 
the BK equation with QCD running coupling (denoted as ``rcBK'' in the 
figure) is used to evaluate the photon spectra in $pp$ and $pPb$ 
collisions at forward rapidities, the ratio $R_{pA}$ is below unity 
in the whole considered range of $p_T$'s, in agreement with 
the results obtained in Ref.~\cite{Ducloue:2017kkq}. Our results 
indicate that a future experimental analysis of the nuclear modification
factor at forward rapidities can be very useful to discriminate between 
these two approaches.
\begin{figure}[t]
\begin{center}
\scalebox{0.65}{\includegraphics{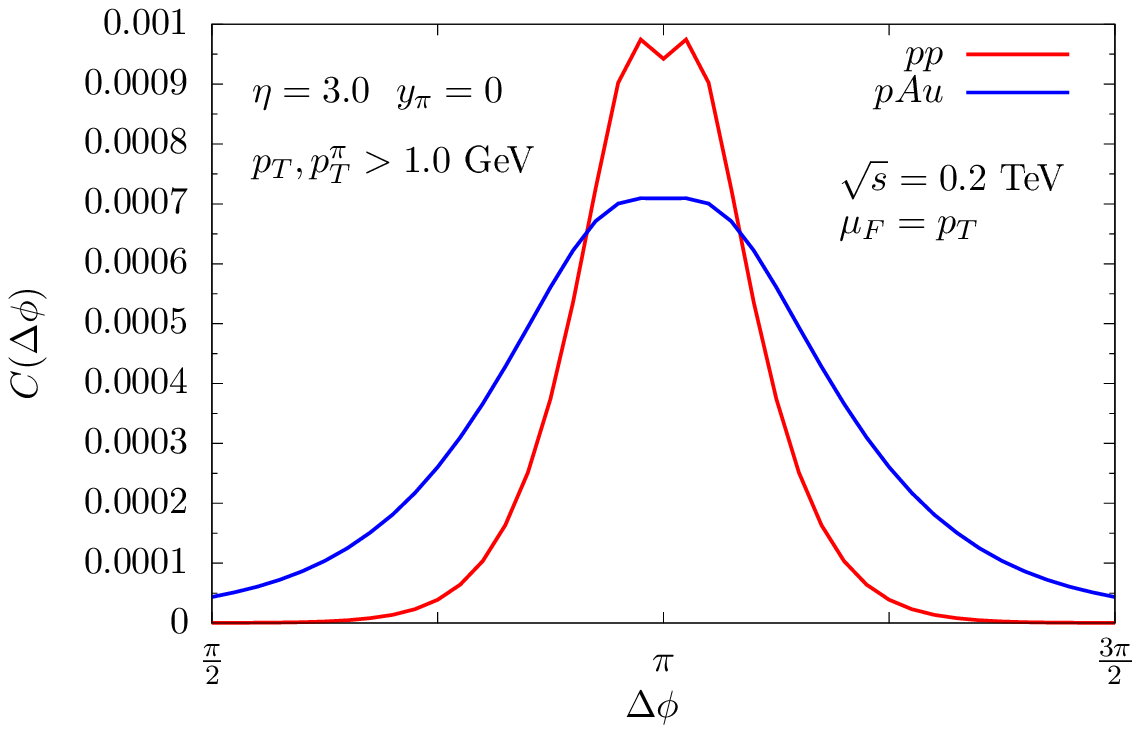}}
\scalebox{0.65}{\includegraphics{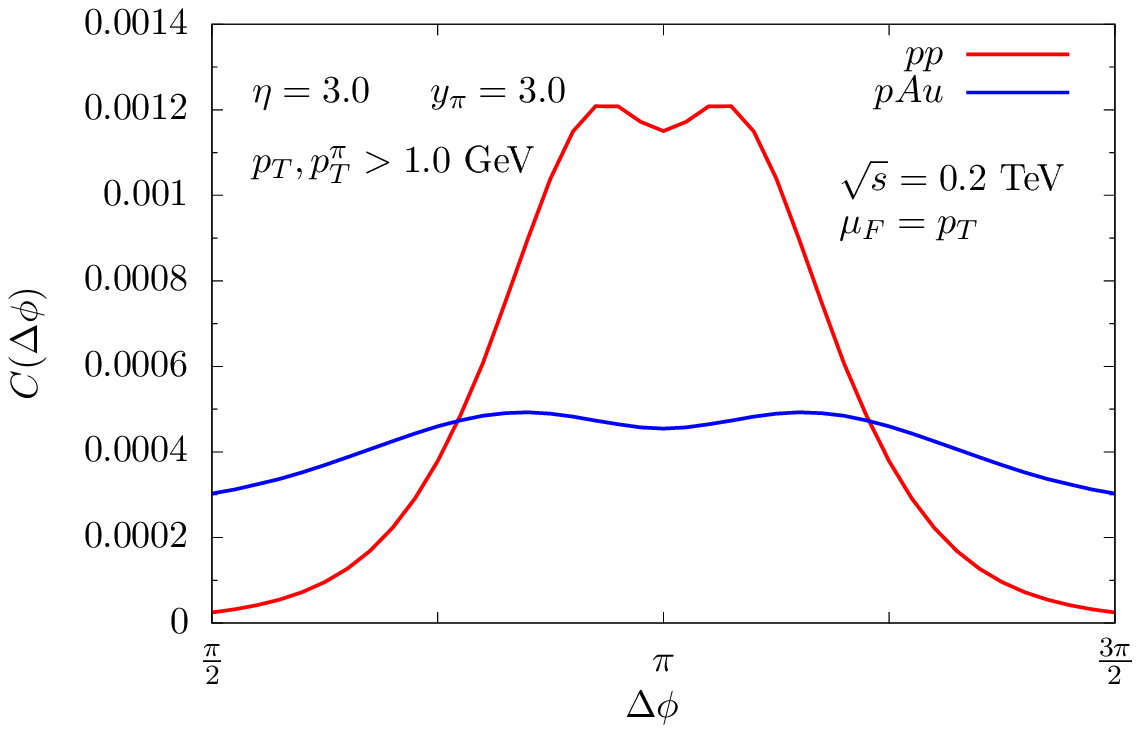}}
\caption{The correlation function $C(\Delta \phi)$ for the associated 
isolated photon and pion production in $pp$ and $pAu$ collisions at RHIC 
($\sqrt{s}=0.2$ TeV) considering two different configurations 
for the photon and pion rapidities.}
\label{fig:cor_rhic}
\end{center}
\end{figure}
\begin{figure}[t]
\begin{center}
\scalebox{0.5}{\includegraphics{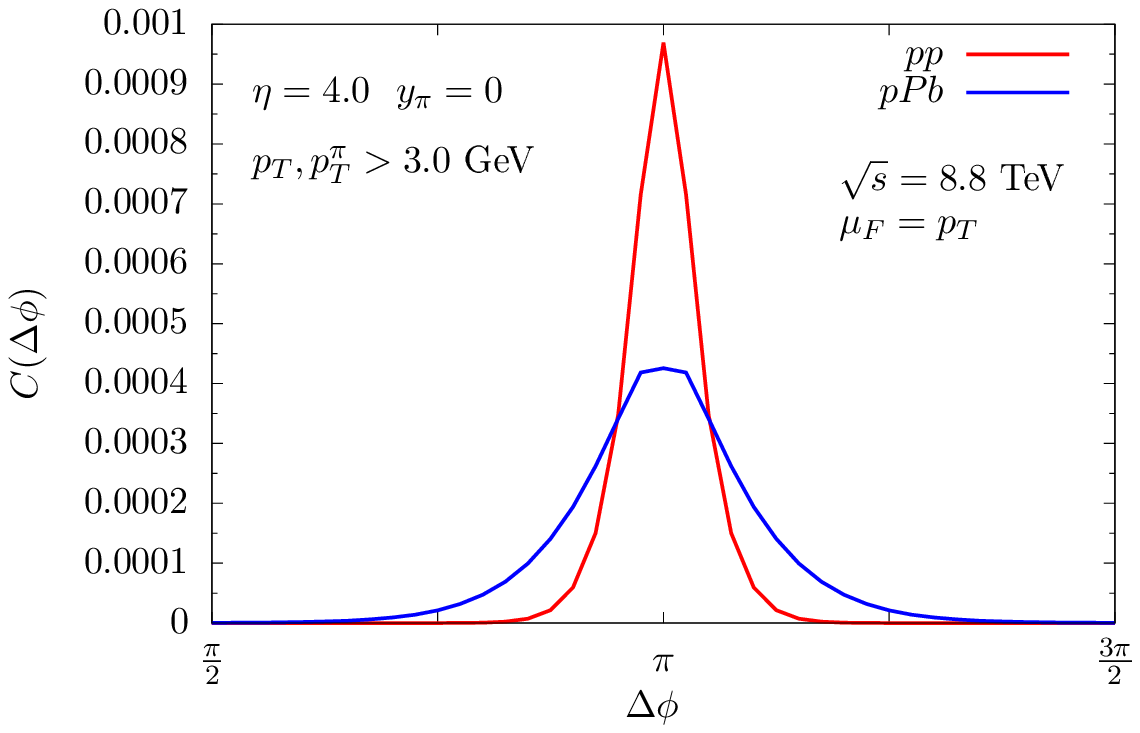}}
\scalebox{0.5}{\includegraphics{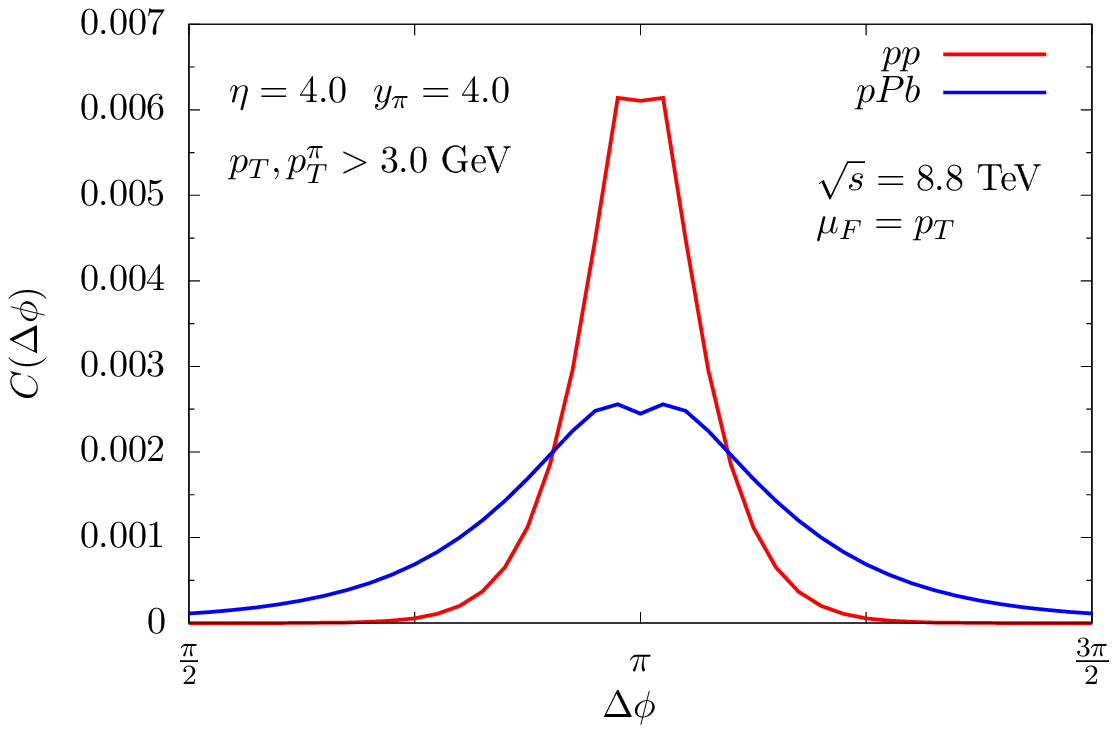}}
\scalebox{0.5}{\includegraphics{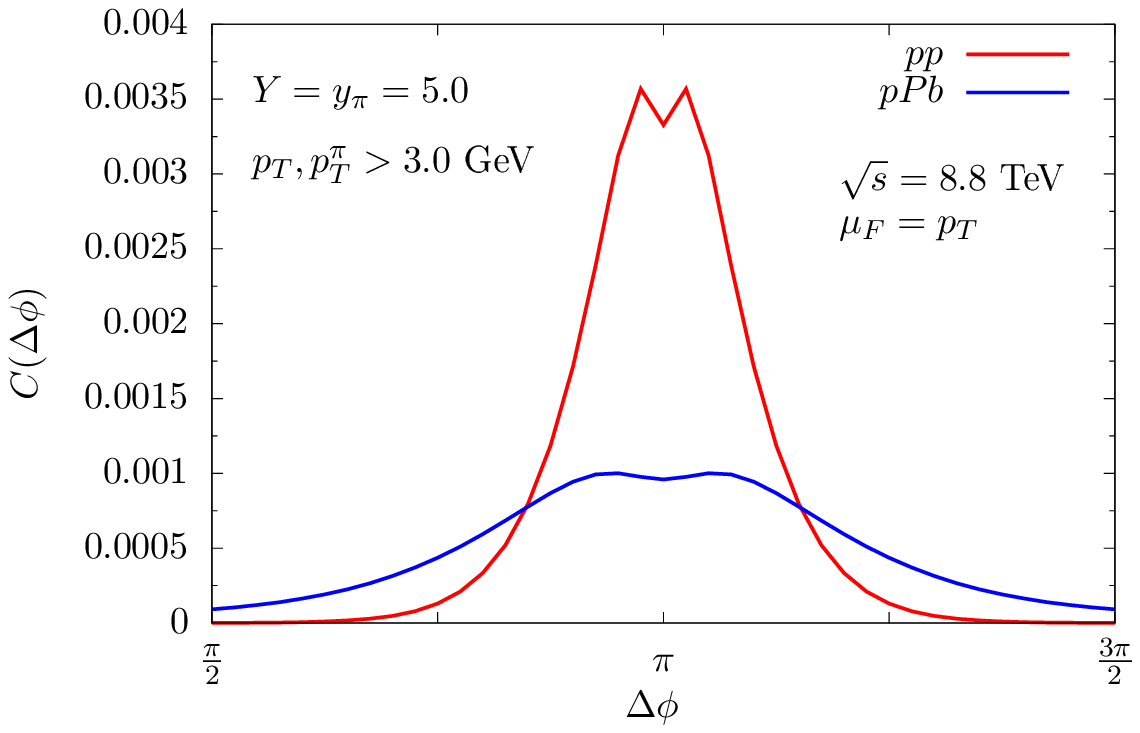}}
\caption{The correlation function $C(\Delta \phi)$ for 
the associated photon and pion production in $pp$ and $pPb$ collisions 
at the LHC ($\sqrt{s}=8.8$ TeV) considering three distinct 
configurations for the isolated photon and pion rapidities.}
\label{fig:cor_lhc}
\end{center}
\end{figure}

In order to probe the underlying dynamics of particle production 
at forward rapidities, one should study other observables sensitive 
to QCD dynamics at small $x$, in particular to QCD non-linear and 
saturation phenomena. An appealing possibility is to consider the
correlation function $C(\Delta \phi)$ defined in Eq.~(\ref{corr}), 
which is strongly sensitive to the details of the dipole model. 
The previous results for the associated DY + pion production 
\cite{Basso:2015pba,BGKNP} have demonstrated that the effect 
of saturation implies a notable smearing of the back-to-back 
scattering profile predicted by the standard collinear formalism.
An addition of the NLO corrections in the collinear framework
would not account for a dip found at $\Delta \phi=\pi$ in the correlation 
function which is a direct manifestation of the saturation phenomenon.

Our goal here is to make the corresponding predictions for the 
isolated photon + pion $h=\pi$ associated production in $pp$ and $pA$ collisions 
at RHIC and LHC energies. As was typically done in earlier studies, 
let us initially consider the GBW model for the dipole cross section 
off the proton target which corresponds to the soft UGDF in the proton
\begin{equation}
F_p(x_g,k^g_T) = \frac{1}{\pi Q_{s,p}^2(x_g) }\, e^{-{k^g_T}^2/Q_{s,p}^2(x_g)} \,,
\label{ung}
\end{equation} 
with the saturation scale given in Eq.~(\ref{satsca}). Following Ref.~\cite{BGKNP}, 
the nuclear UGDF, $F_A$, can also be approximately described by Eq.~(\ref{ung}) 
replacing the proton saturation scale by a nucleus one: 
\begin{eqnarray}
Q_{s,p}^2 \rightarrow Q_{s,A}^2(x) = A^{1/3} c(b)\,Q_{s,p}^2(x) \,,
\end{eqnarray}
where $c=c(b)$ is the profile function of impact parameter $b$ (for central 
collisions, we use $c=0.85$ following Ref.~\cite{stasto}). Moreover, in practical 
calculations we adopt the CT10 NLO parametrization for the parton distributions 
and the Kniehl-Kramer-Potter (KKP) fragmentation function $D_{h/f}(z_h,\mu_F^2)$ 
of a quark into a neutral pion \cite{kkp}. In our analysis, the minimal transverse 
momentum ($p_T^{\rm cut}$) for the photon and the pion in Eq.~(\ref{corr}) 
will be assumed to be the same and equal to 1.0 (3.0) GeV for RHIC (LHC) energies.

In Fig.~\ref{fig:cor_rhic} we present our predictions using the GBW model for 
the correlation function in the case of $pp$ and $pAu$ collisions at RHIC 
($\sqrt{s} = 0.2$ TeV) and for two configurations for the photon and pion rapidities. 
We consider two distinct kinematical configurations, first, when both photon and pion are 
produced at forward rapidities, with $\eta = y_{\pi} = 3$, and, second, when the photon 
is produced at forward rapidity ($\eta = 3$) but the pion is produced at central 
rapidity ($y_{\pi} = 0$). Such configurations can be experimentally studied by 
the STAR Collaboration in both $pp$ and $pA$ collisions. It is important to emphasize 
that the saturation scale increases for smaller values of $x_g$, with 
$x_g = x_1\,e^{-2\eta} +  \frac{x_h}{z_h}\,e^{-2y_h}$, and for larger nuclei. 
Therefore, larger pion and photon rapidities imply the increasing saturation scale. 
Consequently, one should expect a larger decorrelation at forward rapidities and 
at larger values of the atomic mass $A$. 

In addition, for forward rapidities, the transverse momentum of the produced particles 
is limited by the phase space and, in general, does not assume a large value. Therefore, 
for this kinematical range, the saturation scale becomes non-negligible in comparison 
to the typical transverse momentum of the back-to-back scattered particles. In this case, 
the saturation scale induces a noticeable decorrelation between them. Such an expectation 
is confirmed by the results presented in Fig.~\ref{fig:cor_rhic}. For the two configurations 
of rapidities mentioned above, we predict the presence of a double-peak in the correlation function 
in $pp$ collisions with a dip at $\Delta \phi = \pi$, in consistency with the DY + pion analysis of
Ref.~\cite{Basso:2015pba,BGKNP}. Moreover, the width of the double peak increases when 
both rapidities are large. For $pAu$ collisions, the decorrelation grows, with the correlation 
function being almost flat for $\eta = y_{\pi} = 3$. Such a large decorrelation can, in principle, 
be probed in future experimental measurements at RHIC.  
\begin{figure}[t]
\begin{center}
\scalebox{0.75}{\includegraphics{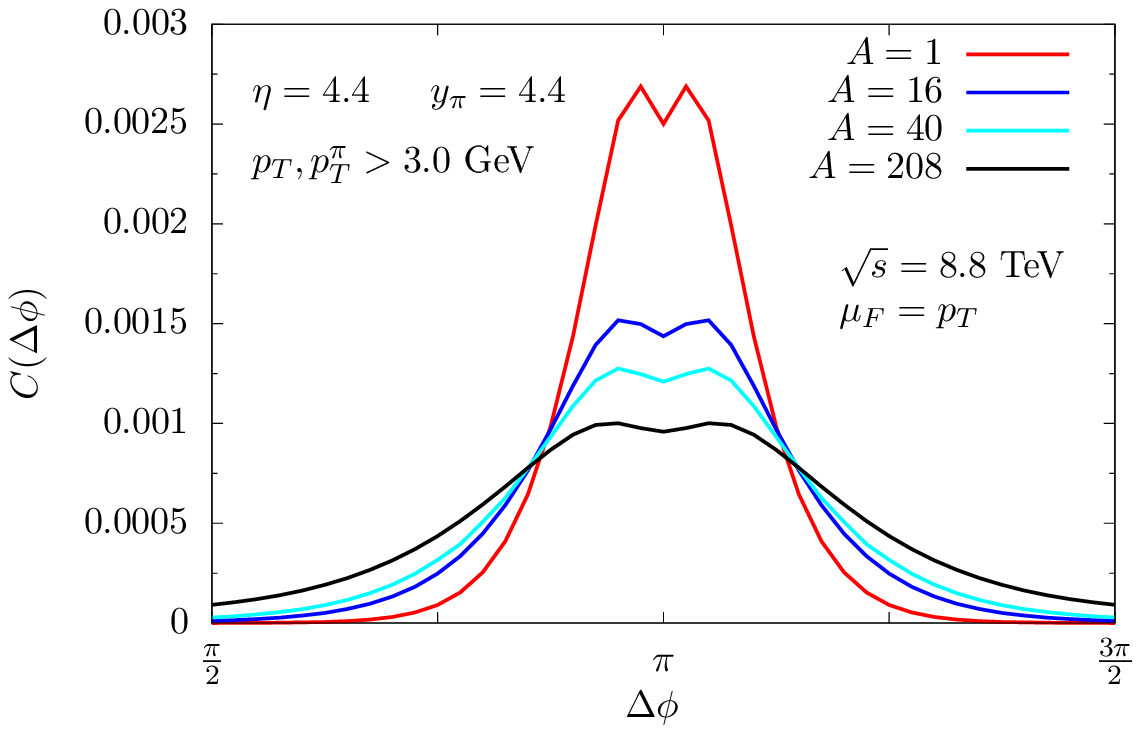}}
\caption{The correlation function $C(\Delta \phi)$ for the associated photon 
and pion production in $pA$ collisions at the LHC ($\sqrt{s}=8.8$ TeV) 
for different nuclei.}
\label{fig:CP_nuclei}
\end{center}
\end{figure}
\begin{figure}[t]
\begin{center}
\scalebox{0.5}{\includegraphics{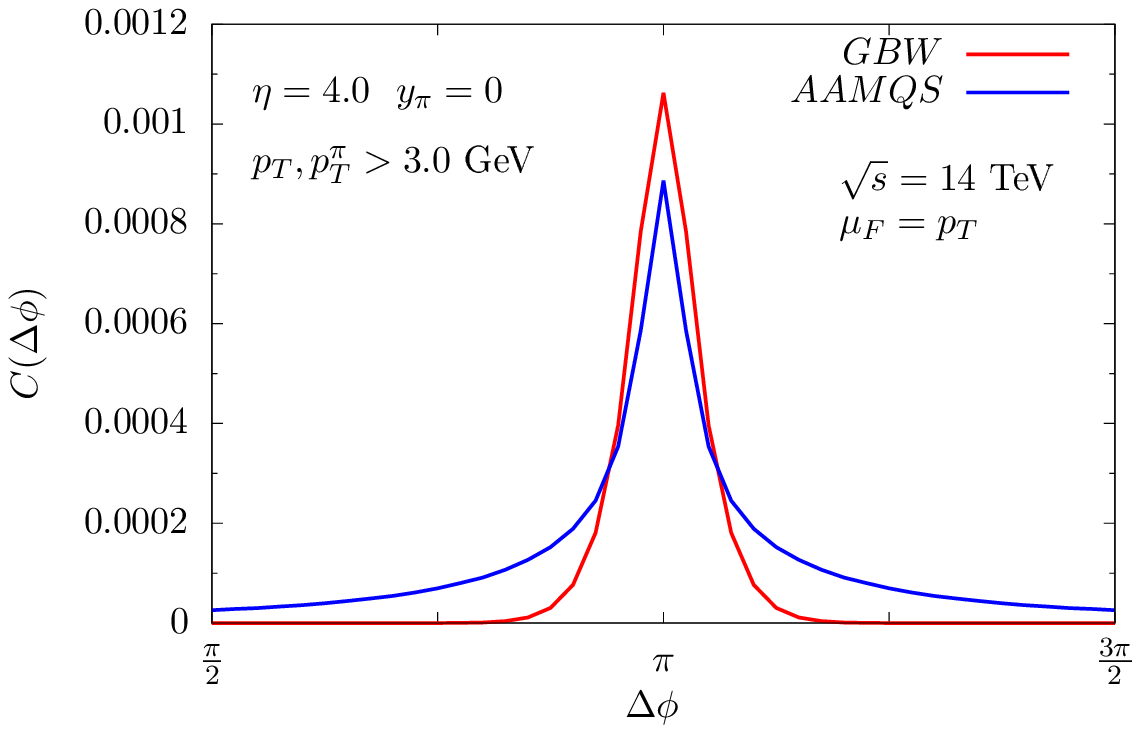}}
\scalebox{0.5}{\includegraphics{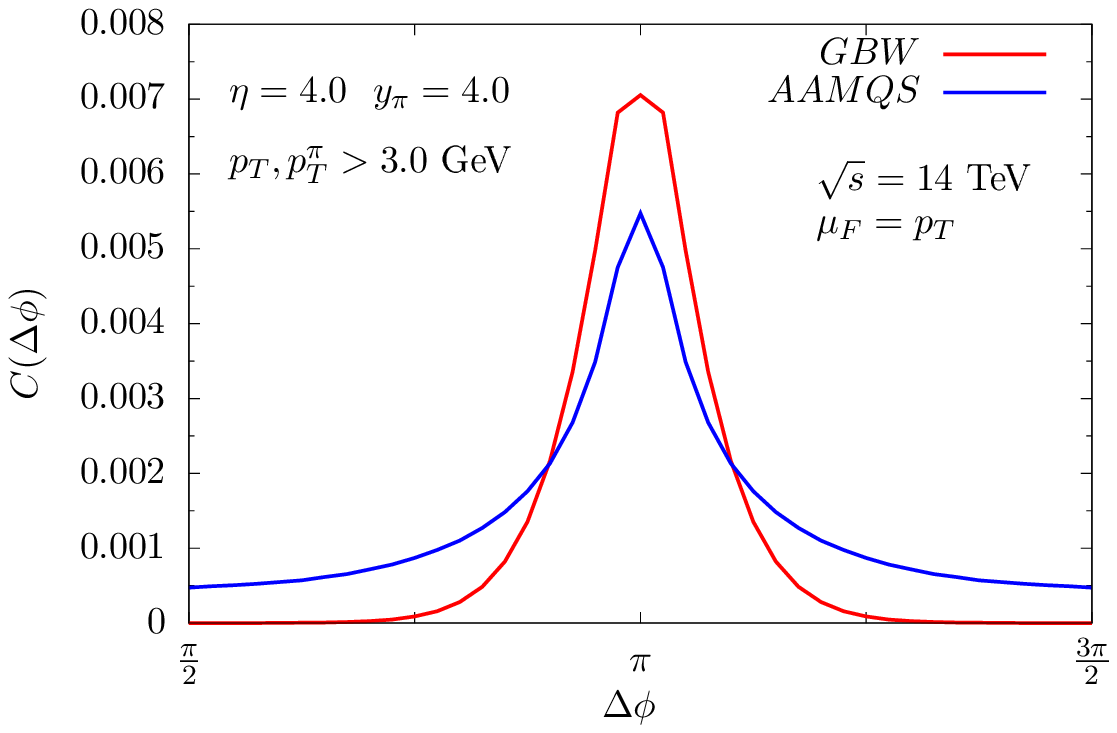}}
\scalebox{0.5}{\includegraphics{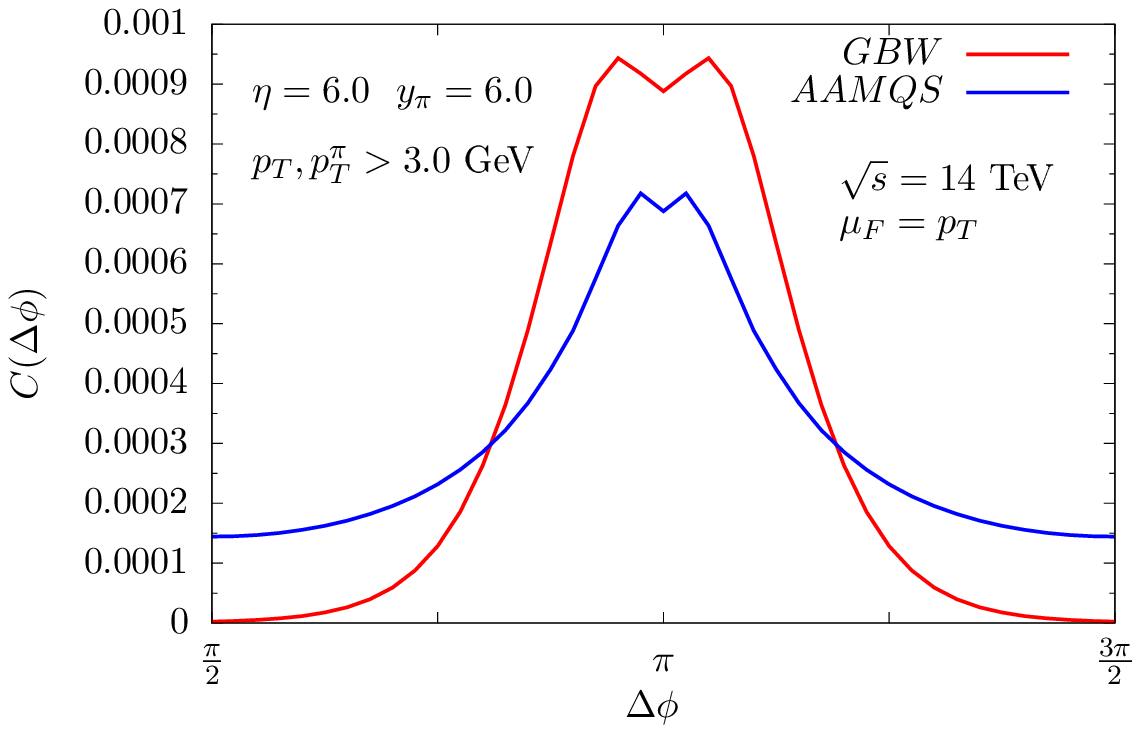}}
\caption{The correlation function $C(\Delta \phi)$ for the associated photon 
and pion production in $pp$ collisions at the LHC ($\sqrt{s}=14$ TeV) considering 
three configurations for the photon and pion rapidities and the GBW and AAMQS models 
for the UGDF in the proton target.}
\label{fig:models}
\end{center}
\end{figure}

Our predictions for the correlation function in $pp$ and $pPb$ collisions at $\sqrt{s} = 8.8$ TeV
are presented in Fig.~\ref{fig:cor_lhc}. In the case of $pp$ collisions, we notice a smearing of the
back-to-back correlation when the rapidities are increased, which is directly related to the 
growth of the saturation scale. A similar behavior is predicted for the DY + pion process ~\cite{Basso:2015pba,BGKNP}. 
In contrast, for $pPb$ collisions at the same center-of-mass energy, we predict a larger 
decorrelation, in particular for $\eta = y_{\pi} = 5$. 

In order to analyze the impact of the atomic mass on the correlation function, 
in Fig.~\ref{fig:CP_nuclei} we present our predictions for $C(\Delta \phi)$ in 
$pA$ collisions at $\sqrt{s} = 8.8$ TeV, different values of $A$ and $\eta = y_{\pi} = 4.4$. 
As expected due to a growth of the saturation scale with $A$, we observe that the decorrelation 
becomes stronger for heavier nuclei. Such a result indicates that, in principle, the study 
of $C(\Delta \phi)$ for a fixed energy and for given set of rapidities can be used to probe 
the $A$-dependence of the nuclear saturation scale $Q_{s,A}$.

Finally, let us discuss how the above predictions for the correlation function depend on modelling 
of the UGDF in the target. For the latter, so far we have used Eq.~(\ref{ung}) inspired by the GBW 
model as input in our calculations. It is interesting to compare these results with those obtained by 
using the solution of the rcBK equation discussed above. In Fig.~\ref{fig:models} we present 
a comparison between the GBW predictions and those derived using the AAMQS model. 
We observe that both models predict a similar behaviour for the correlation function and 
differ mainly in the near-side ($\Delta \phi = 0$) region, which is dominated by the leading 
jet fragmentation. Such a result is anticipated from the previous studies 
\cite{stasto,Stasto:2018rci,amir} which have demonstrated that the behaviour of $C(\Delta \phi)$ 
in the away-side ($\Delta \phi = \pi$) region is not strongly dependent on the large transverse 
momentum tail of the UGDF.

{
Recently, the formalism of resummation of the Sudakov-type double logarithms at small-$x$ 
has been developed \cite{yuan1,yuan2}. Such terms appear in the description of the transverse-momentum 
spectrum of a given hard process due to presence of two scales. In the considered process, these two scales are 
the total transverse-momentum imbalance of the $\gamma + \pi$ system, $q_T = |\vec{p}_T + \vec{p}_T^{\,\,\,\pi}|$, 
and the average transverse momentum $ Q_T = |\vec{p}_T - \vec{p}_T^{\,\,\,\pi}|/2 $. In the kinematical domain of 
a large scale separation, $Q_T \gg q_T$, large logarithms of type $\ln^2 Q_T^2/q_T^2$ appear in every order 
of perturbative calculations and need to be resummed. 
As demonstrated in Ref.~\cite{yuan2} for jet-photon production in $pA$ collisions when considering the corrections 
associated to single gluon radiation, the contribution of the Sudakov double logarithms can be factorized from 
the small-$x$ logarithms provided that these two contributions are well separated in the phase space of the 
radiated gluon. This formalism was used in Ref.~\cite{Stasto:2018rci} to estimate the dihadron angular correlations 
in forward $pA$ collisions, which demonstrated that the Sudakov correction becomes important in some specific regions 
of the phase space.  
One important question is the impact of the Sudakov corrections on the dip structure observed in our results. As demonstrated 
above, the dip appears when the typical transverse momenta of the photon and pion are relatively close to the saturation scale. 
Moreover, we have noticed that the total transverse momentum imbalance $q_T$ is of the order of $Q_s$ indicating that the dip 
arises when $Q_T \approx q_T$. In this regime, the Sudakov logarithms $\ln^n Q_T^2/q_T^2$ are typically small. Consequently, 
we do not expect any significant effect on the angular distributions and, in particular, on the dip structure associated to the Sudakov 
resummation in the relevant kinematic domains. Surely, the inclusion of such small corrections can be considered for a future work, 
with the results presented here being the starting point.
}

\section{Summary}
\label{Sec:summary}

In this paper, we performed a detailed phenomenological analysis of the isolated photon production in $pp$ and $pA$ collisions at typical 
RHIC and LHC energies in the framework of color dipole approach. We employed three different phenomenological saturation models 
for the dipole-target scattering and analysed differential distributions of prompt photons in transverse momentum $p_T$. Besides, we have investigated 
the correlation function $C(\Delta \phi)$ in azimuthal angle between the real high-$p_T$ photon produced in association with a leading pion emerging via
fragmentation of a projectile quark which emits the photon. This observable has been studied in $pp$ and $pA$ collisions at RHIC and LHC energies
and at different rapidities of final states. In $pp$ collisions, the correlation function exhibits a double-peak structure close to $\Delta \phi \simeq \pi$
in certain kinematical configurations corresponding to both the real high-$p_T$ photon and the accompanied high-$p_T$ pion being produced 
at forward rapidities. In the case of $pA$ collisions, a larger nuclear saturation scale enforces a stronger decorrelation between the photon and the pion.
The correlation function is a more exclusive observable than the standard transverse momentum spectra of the isolated photon and appears to be strongly sensitive
to the details of theoretical modelling of the saturation phenomena in QCD. A future measurement of this observable at different RHIC and the LHC energies 
would be capable of setting stronger constraints on the unintegrated gluon density in the small-$x$ and small-$k_T$ domains as well as 
on the dipole model parametrizations, thus, enabling to directly probe the saturation scale.

\section*{Acknowledgements}

V. P. G. thanks the members of the Faculty of Nuclear Sciences and Physical Engineering of the Czech Technical University in Prague 
and of the Nuclear Physics Institute of the CAS for warm hospitality during the completion of this work. V. P. G. and Y. L. were  partially financed 
by the Brazilian funding agencies CNPq, CAPES, FAPERGS and INCT-FNA (process number 464898/2014-5).
M. \v{S}. is partially supported by the grants LTT17018 and LTT18002 of the Ministry of Education of the Czech Republic and by the grant 13-20841S 
of the Czech Science Foundation (GACR).
R. P. is partially supported 
by the Swedish Research Council grant No. 2016-05996, 
by the European Research Council (ERC) under the European Union's Horizon 2020 
research and innovation programme (grant agreement No 668679), as well as by 
the Ministry of Education, Youth and Sports of the Czech Republic project LTT17018 
and by the NKFI grant K133046 (Hungary).
Part of this work has been performed in the framework of COST Action CA15213 “Theory
of hot matter and relativistic heavy-ion collisions” (THOR).

\bibliographystyle{unsrt}

\end{document}